\documentclass[conference]{IEEEtran}

\usepackage{amsmath} 
\usepackage{graphicx}
\usepackage{algorithm}
\usepackage{algorithmic}

\makeatletter
\def\blfootnote{\gdef\@thefnmark{}\@footnotetext}
\makeatother

\usepackage{bm}
\usepackage{multirow}

\newcommand{\authormarkone}{$^\dagger$}

\if
\usepackage{CJKutf8}
\def\ToDo#1{\begin{CJK}{UTF8}{ipxg}
	     \textbf{ToDo: #1}
	    \end{CJK}}
\def\ToTranslate#1{\begin{CJK}{UTF8}{ipxm}
		    \textbf{Translate:} #1
		   \end{CJK}}
\fi
\def\ToDo#1{}
\def\ToTranslate#1{}

\begin{document}

\title{Mercem: Method Name Recommendation\\Based on Call Graph Embedding}




\author{
    \IEEEauthorblockN{Hiroshi Yonai \authormarkone}
    \IEEEauthorblockA{
        Graduate School of\\
        System and Information Engineering,\\
        University of Tsukuba\\
        Tsukuba, Japan\\
        yonai@kde.cs.tsukuba.ac.jp
    }
    \and
    \IEEEauthorblockN{Yasuhiro Hayase \authormarkone}
    \IEEEauthorblockA{
        Faculty of\\
      	Engineering, Information and Systems,\\
        University of Tsukuba\\
        Tsukuba, Japan\\
        hayase@cs.tsukuba.ac.jp
    }
    \and
    \IEEEauthorblockN{Hiroyuki Kitagawa}
    \IEEEauthorblockA{
        Center for Computational Sciences,\\
        University of Tsukuba\\
        Tsukuba, Japan\\
        kitagawa@cs.tsukuba.ac.jp
    }
}
\maketitle
\blfootnote{\authormarkone The first two authors contributed equally to this work}

\maketitle

\begin{abstract}
  Comprehensibility of source code is strongly affected by identifier names, therefore software developers need to give good (e.g. meaningful but short) names to identifiers.
  On the other hand, giving a good name is sometimes a difficult and time-consuming task even for experienced developers.
  To support naming identifiers, several techniques for recommending identifier name candidates have been proposed.
  These techniques, however, still have challenges on the goodness of suggested candidates and limitations on applicable situations.
  This paper proposes a new approach to recommending method names by applying graph embedding techniques to the method call graph.
  The evaluation experiment confirms that the proposed technique can suggest more appropriate method name candidates in difficult situations than the state of the art approach.
\end{abstract}


\section{Introduction}
Program comprehension is essential activity through a lifecycle of software development. \cite{von1995program}
This is because, for each change request, a software developer decides what parts of a program should be changed to accomplish a requirement, and to assess what parts of the program are impacted by the change.
This exploration is necessary even for a tiny change request.
In an agile software development project, a software product is incrementally changed day by day, so that the source code is strongly required to be easy to comprehend. \cite{beck2001agile}
Also in case of waterfall project, program comprehension is a costly activity, since most part of the costs in a maintenance phase is caused by program comprehension\cite{murphy2005emergent}, and the maintenance cost sometimes occupies 70\% of lifecycle cost of the software\cite{lientz1978characteristics}.

To help program comprehension, an identifier name is desired to be a clue to guess the role and behavior of a corresponding program entity.
Contrary to expectation, identifiers can have obscure or meaningless names, and the time required to comprehend a program is increased, as well as the correctness of understanding is impaired. \cite{lawrie06whats}
Consequently, identifiers should have \textit{good} names, namely, meaningful, distinct, specific, moderately short, and understandable for average developers. \cite{martin2008clean-chap2,boswell2012art-chap2}
However, developers sometimes feel difficulty in naming, because good naming requires not only exhaustive understanding of the target entity based on knowledge of the target domain of a developed program, but also the custom of identifier naming, including usage of the terms, that is only learned from the experience of software development. \cite{martin2008clean-chap2}

To alleviate the difficulty of identifier naming, several techniques are proposed for recommending names for a method in OOP programs.
Kashiwabara et al. proposed an approach to recommend a verb part of a method name by leveraging association rules, learned from a source code corpus, between a verb in a method name and elements contained in a corresponding method body. \cite{kashiwabara2014recommending}
Allamanis et al. proposed an approach for recommending identifier names using \textit{word-embedding} for identifier names. \cite{allamanis2015suggesting}
This approach extract skip-gram embedding\cite{mikolov2013linguistic} of identifiers by treating source code in the same way as natural language text.
Since this approach requires numerous surrounding contexts of appearances of a target identifier,
a meaningful recommendation for a bad identifier name is unavailable before the bad name is widely used.
Allamanis et al. also proposed another approach to estimate a method name by summarizing a method body.\cite{allamanis2016convolutional}
This summarization-based approach achieves high precision when words in the true method name appear in the method body, or a frequent programming idiom strongly correlates with the words in a method name, like a getter or setter.
However, when keywords of a method name are contained in the body, or when a method is boilerplate, naming the method tends to be easy for developers.

In this paper, we propose \textit{Mercem} (MEthod name Recommendation based on Call graph EMbedding), a novel approach to recommend a method name for developers struggling with naming a method.
This approach can output good recommendations for a method, which is not a getter or a setter and whose body does not contain words in the true method name before the method is called.
The key idea of the proposed approach is to recommend names of methods whose function is similar to a target (i.e. a method to be renamed) in terms of a method embedding that expresses the function of a method, obtained from \textit{caller-callee} relationships.
In the proposed approach, the recommendation is available immediately after the method body is created and, of course, before the method is called since the approach only requires a set of methods, which the target method calls.

An evaluation experiment is performed to compare the proposed approach and the state of the art\cite{allamanis2016convolutional}.
The purpose of the experiment is to determine which the proposed technique and the state of the art outperform the other technique in the difficult situation, i.e. where a target method is not a getter or a setter and a word used in the correct method name is unavailable in the method body.
Additionally, factors in the results of the evaluation are investigated exploratorily.

The remainder of this paper is structured as follows:
Section 2 introduces the basic knowledge and related work;
Section 3 explains details of the proposed approach;
Section 4 shows an evaluation experiment and its results;
Section 5 discusses the result of the experiments;
And finally, section 6 describes conclusion and future remarks.

\section{Related work}

\subsection{Distributed Representation and Embedding}
Distributed representation is a way to represent an entity as a row of real numbers, i.e. a vector.
Distributed representation was initially proposed as a technique for treating multi-type data in neural networks\cite{Hinton:1986:DR:104279.104287} but is widely applied in many fields nowadays.
In research fields of natural language processing, distributed representation plays a crucial role in alleviating the sparsity of words, and is usually referred to as \textit{embedding}.
Additionally, \textit{word2vec}\cite{mikolov2013linguistic,mikolov2013efficient,mikolov2013distributed} succeeded in relating the relative position between embedding vectors and the semantic relationship of words, leading to numerous applications and derivative studies. Word2vec is based on the distribution hypothesis\cite{harris1954distributional}; that is the hypothesis that there is a relationship between the meaning of a certain word and surrounding words (i.e. context) of the appearances of the word.

Based on the success in NLP, distributed representations for elements of graphs are actively studied and recognized as an effective yet efficient way to solve the graph analytics problem\cite{DBLP:JOURNALS/CORR/ABS-1709-07604}.
Distributed representations are calculated not only for nodes or edges, but also for subgraphs or entire graphs, but in the following, we will only focus on node embeddings used in the proposed technique.
There are three representative types of proximity measures, which is the graph properties preserved in the embeddings.
\textit{First-order proximity} is a concept that directly connected two nodes corresponding to close vectors.
\textit{Second-order proximity} is the concept that two nodes that have similar sets of connected nodes corresponding to close vectors.
\textit{Higher-order proximity} is a concept that two nodes whose neighborhood including nodes distant 2 or more apart are similar correspond to close vectors.
As a variant of the higher-order proximity technique, random walk based node embedding techniques\cite{Perozzi:2014:DOL:2623330.2623732,Grover:2016:NSF:2939672.2939754}, which apply the word embedding technique to sequences of nodes generated by random walking on the input graph, have performed well in node classification and clustering tasks.

\subsection{Recommending Identifier Names}
For solving the difficulty of naming identifiers in the source code, several approaches are proposed to evaluate or recommend identifier names.

H{\o}st et al.\cite{host2007programmer} found dozens of rules between a verb part of method name and its body.
Leveraging these rules, they also proposed a technique to automatically evaluate whether the verb in a method name is appropriate for its body.\cite{host2009debugging}
These approaches are difficult to be scaled to other verbs since the rules are acquired manually.

Kashiwabara et al. \cite{kashiwabara2014recommending}
proposed a technique to recommend verb part of a method name leveraging association rules between a method name and related information, such as local variable names, parameter names, return type, class names and so on.
Comparing to H{\o}st's approach\cite{host2009debugging}, this technique can support a variety of verbs since the rules are automatically built from a code corpus.

Allamanis et al. \cite{allamanis2015suggesting} proposed a technique to recommend identifier names using word-embedding.
This approach treats source code as a sequence of tokens, and then calculate the skip-gram embedding\cite{mikolov2013linguistic} from the sequence.
Due to employing skip-gram technique, this approach requires numerous appearances of an identifier for a precise recommendation, since enough surrounding contexts of the appearances are needed to obtain meaningful embedding of the identifier.
This means that meaningful recommendation for a method or class is unavailable before the method/class is used enough.

Allamanis et al.\cite{allamanis2016convolutional} also proposed another approach to estimate a method name based on a neural summarization technique.
The body of a target method is separated into a sequence of words, (namely, all identifiers are split into discrete words), then the sequence of words is input to the summarization system.
This summarization-based approach achieves high precision when words in the true method name appear in the method body, or a frequent programming idiom strongly correlates the words in a method name, like a getter or setter.
However, when keywords of method name are contained in the body, or when a method is only composed of a few frequent idioms, naming the method tends to be easy for developers, so the demand for supporting identifier naming could be relatively low.
Note that this approach does not explicitly take into account the function or role of a class or a method appearing in a target method since all identifiers are split into discrete words before input to the summarization system.






\section{Proposed Approach} \label{sec:methodology}

\begin{figure*}[tbp]
  \centerline{\includegraphics[width=2\columnwidth,pagebox=cropbox,clip]{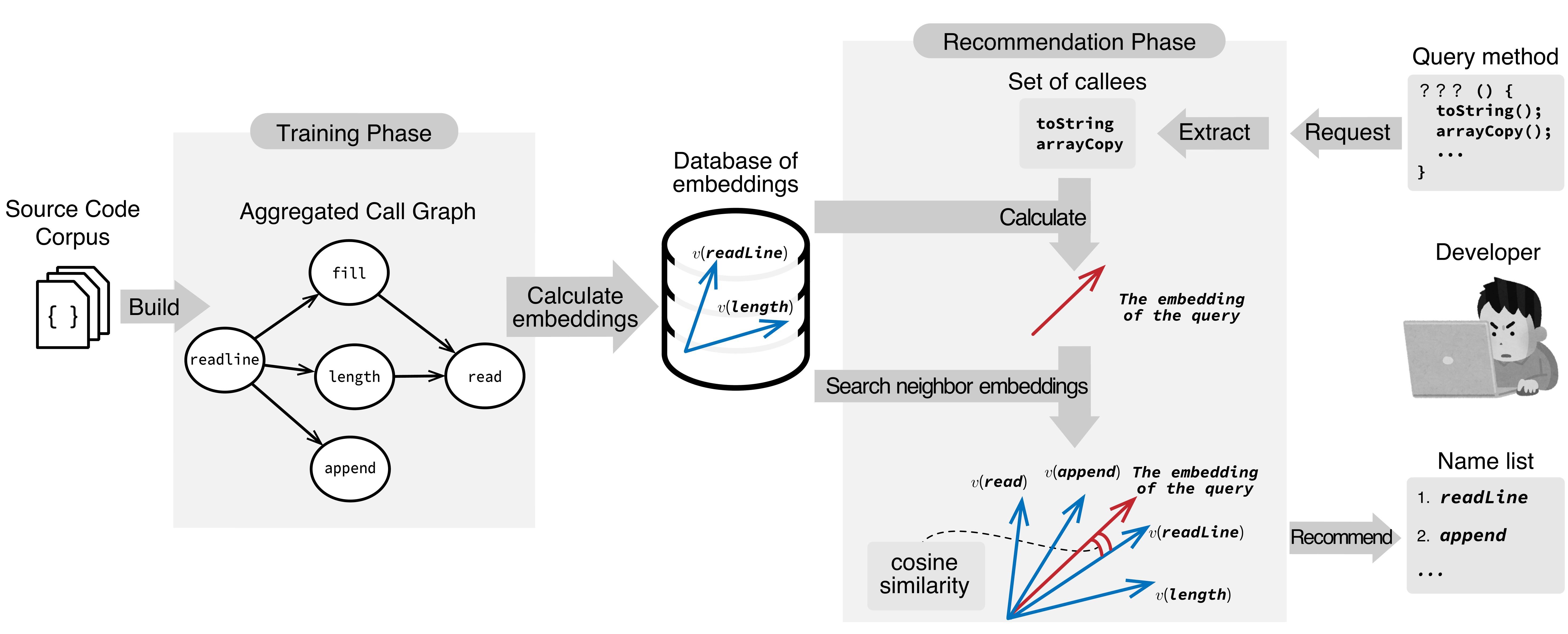}}
  \caption{Overview of Mercem}
  \label{fig:overview}
\end{figure*}

This section describes the details of the proposed approach Mercem.
The key idea of the approach is to recommend names of existing methods whose function is similar to the target (i.e. a method to be renamed) in terms of method embeddings calculated from caller-callee relationships.
Fig. \ref{fig:overview} shows an overview of the proposed approach.
The approach consists of two phases: \textit{training phase} to be run once in advance, and \textit{recommendation phase} invoked by a developer on demand.
The Role of the training phase is to calculate embeddings of method names appearing in a code corpus, and to store the embedding in a database preparing for recommendations.
The recommendation phase provides a list of candidate names for a target method to a user using the method embeddings built by the training phase.

In order to associate the function of a method with its embedding, we leverage the relationships that methods use other methods to implement its function.
In other words, the embedding of a method should be placed near to the embedding of a callee,
since the caller method includes functions of the callee method.
Applying this policy to all callees of a method, the embedding of the method will approximate to the average of embeddings of the callees.
This policy for placement can be interpreted as a variant of second-order proximity so that two methods having similar sets of callees will have close embeddings.


In order to find the names of methods whose functions are similar to the query method provided by a user, the recommendation phase calculates the embedding of the query method and then searches similar embeddings from the database.
Note that the embedding of the query method must satisfy all of the following constraints:
\begin{enumerate}
 \item The contexts of invocations of the query method cannot be used, since recommendation must be available before the method is used. \label{item:constraint-on-context}
 \item The time for calculating an embedding of a query method is at most a few seconds, since recommendation results must be available immediately in actual usage scenarios.
       \label{item:constraint-on-time}
 \item 
       Embeddings of query methods must be semantically consistent with the embeddings stored in the database. \label{item:constraint-on-consistency}
\end{enumerate}

To satisfy these constraints, the recommendation subsystem looks up the embedding of each callee of the query method from the database and then searches the nearest neighborhoods of the arithmetic mean of these embeddings in the database.
Constraint 1 is satisfied because the input is obtained only from the body of the query.
Constraint 2 is also satisfied because it only requires picking up callee names from the target method and looking up the embeddings for the names from the database.
As to the consistency, the concept of averaging the callees is the same, and the embeddings in the database are used for the average, so that Constraint 3 is also satisfied.


Details of the training phase and the recommendation phase are explained below.

\subsection{Training phase} \label{subsec:build_call_graph}
The training phase calculates the embeddings of methods by optimization to approximate embedding of each method and the average of embeddings of callees of the method each other, based on the above mentioned policy.


\begin{figure}[tbp]
    \centerline{\includegraphics[width=1.0\columnwidth,pagebox=cropbox,clip]{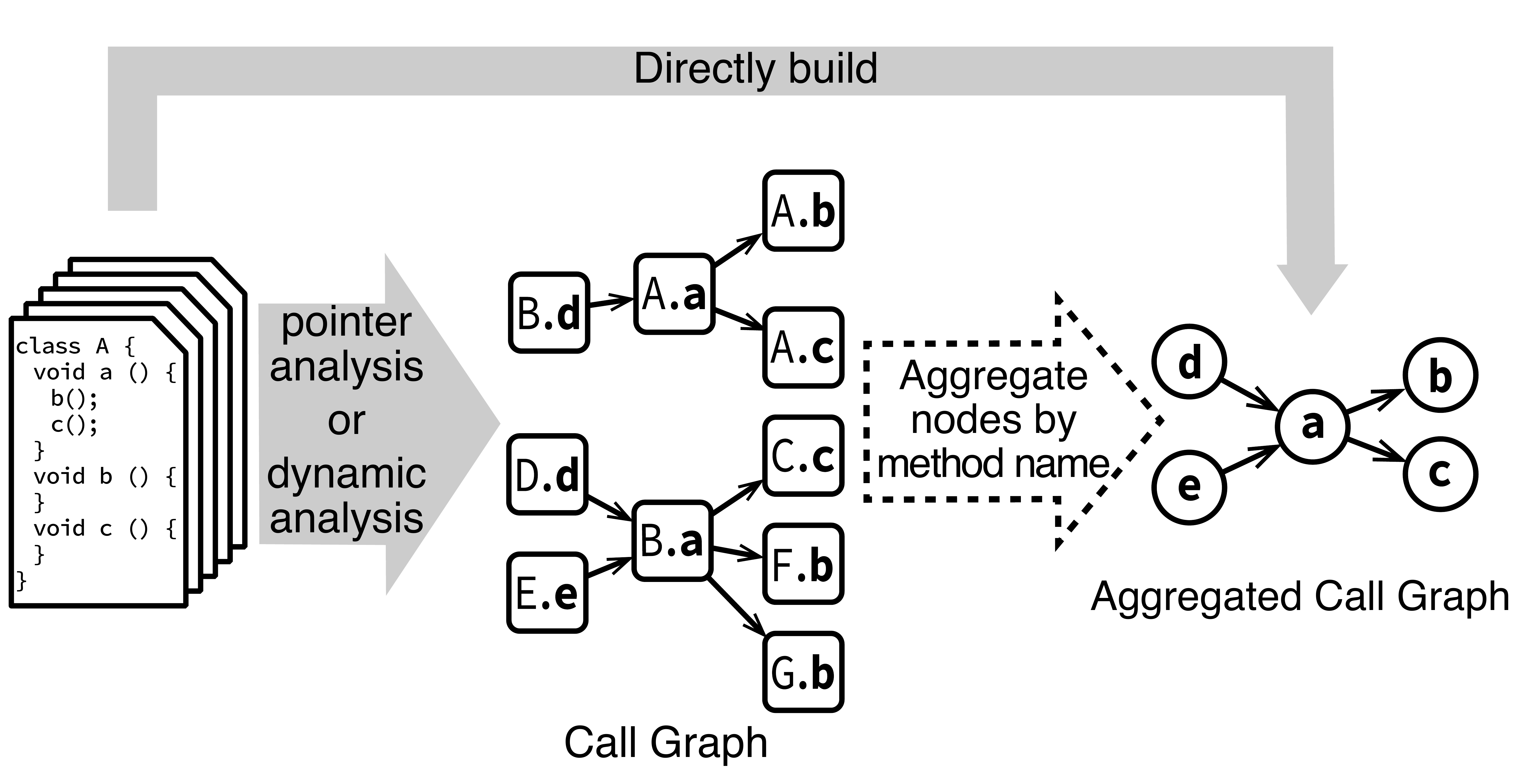}}
    \caption{Example of Aggregated Call Graph}
    \label{fig:concept_of_acg}
\end{figure}

In order to cope with incompleteness of the source code corpus and to reduce the computational cost, the \textit{Aggregated Call Graph} (\textbf{ACG}) is used instead of a call graph. (Fig. \ref{fig:concept_of_acg})
The ACG is a directed graph similar to a call graph, but one node represents one set of methods with the same name, and a directed edge means there are one or more calls between two method sets.
In order to build a precise call graph of a program written in an object-oriented language such as Java, dynamic analysis\cite{Ryder:1979:CCG:1313331.1313683} or pointer analysis\cite{Liang:2001:EEF:379605.379676,Smaragdakis:2015:PA:2802194.2802195} must be applied to a complete set of source or binary code.
However, a common code corpus is incomplete because it sometimes lacks necessary libraries for execution and/or contains uncompilable source files, which may be incorrect or requires special preprocessing.  Therefore manual treatment is required to obtain a complete code set.
Even if a complete code is obtained, pointer analysis or dynamic analysis are costly processes in terms of time and space, so that scaling the corpus is difficult.
Furthermore, when analyzing multiple software products at the same time, two or more methods may have the same fully qualified name, because different versions of libraries are used. In this case, the complete call graph can not exist.
On the other hand, the time to calculate the node embeddings increases with the number of nodes and edges of the graph, because the calculation is based on iterative optimization on the graph.
ACG is designed to overcome these problems; obtaining caller-callee relationships in the ACG is simplified by identifying nodes only by method names, and the computational cost for calculating embeddings on the ACG is reduced by decreasing the number of nodes and edges.
Even if nodes are aggregated in this way, embeddings of method names can be calculated because the relationships between method names are preserved.

Details of building an ACG and calculating embeddings are described below.



\subsubsection{Building an aggregated call graph}

In this step, the ACG is built from source code.
As mentioned above, although the ACG is the same as a call graph whose nodes are grouped by method names, the ACG is not built via the call graph, but directly from source code to avoid the difficulty of constructing a call graph.
The algorithm to calculate an ACG is shown in Algorithm \ref{alg:calc-agc}.
From each of the method definitions extracted from the code corpus, obtain the name of the method and the set of all method names contained in the body. Then, add all those method names to the vertex set $V_A$ of ACG.  Next, edges from the name of the definition to each method name in the body are added to the edge set $V_A$ of the ACG.
This calculation does not require semantic analysis, so the computational cost is low.  Additionally, since each source file is processed independently, the process can be parallelized easily.

\begin{algorithm}
\caption{Calculate ACG}
\label{alg:calc-agc}
\begin{algorithmic}
\STATE $V_A \leftarrow \emptyset$
\STATE $E_A \leftarrow \emptyset$
\FORALL{$\text{mdef} \in \text{all\_method\_definitions}$}
   \STATE $V_A \leftarrow V_A \cup \{\text{mdef.name}\}$
   \FORALL{$\text{iexp} \in \text{mdef.invocation\_expressions}$}
     \STATE $V_A \leftarrow V_A \cup \{\text{iexp.method\_name}\}$
     \STATE $E_A \leftarrow E_A \cup \{(\text{mdef.name}, \text{iexp.method\_name})\}$
   \ENDFOR
\ENDFOR
\RETURN $G_A = (V_A, E_A)$
\end{algorithmic}
\end{algorithm}


\subsubsection{Calculating method embeddings from aggregated call graph}

This step computes the embeddings of method names from ACG.
An embedding of a method name should be placed close to the arithmetic mean of embeddings of the callee names.
At the same time, method names that are not directly connected on the ACG should be placed far apart.

In order to satisfy these requirements, stochastic gradient descent with negative sampling is applied for a loss function, which expresses the requirements.
The loss function $L$ is shown below.
\begin{equation}
 \begin{split}
   L(M) & =
 \alpha \sum_{m \in M} |v(m) - \frac{1}{|C(m)|}\sum_{c \in C(m)} v(c)|^2 \\
  & + (1 - \alpha) \sum_{m \in M} (1 - |v(m)|)^2
 \end{split}
\end{equation}
where $M$ is a set of method names, $v(m)$ is the embedding of the method name $m$, $|v(m)|$ is the L2 norm of $v(m)$, $C(m)$ is a set of the callees of $m$, and $\alpha$ is a weighting parameter.
The first term increases when each embedding is away from the average of embeddings of its callees.
The second term increases when norms of the embeddings divert from 1, in order to prevent the embedding from becoming extremely large or small.

\begin{figure}[tbp]
    \centerline{\includegraphics[width=1.0\columnwidth,pagebox=cropbox,clip]{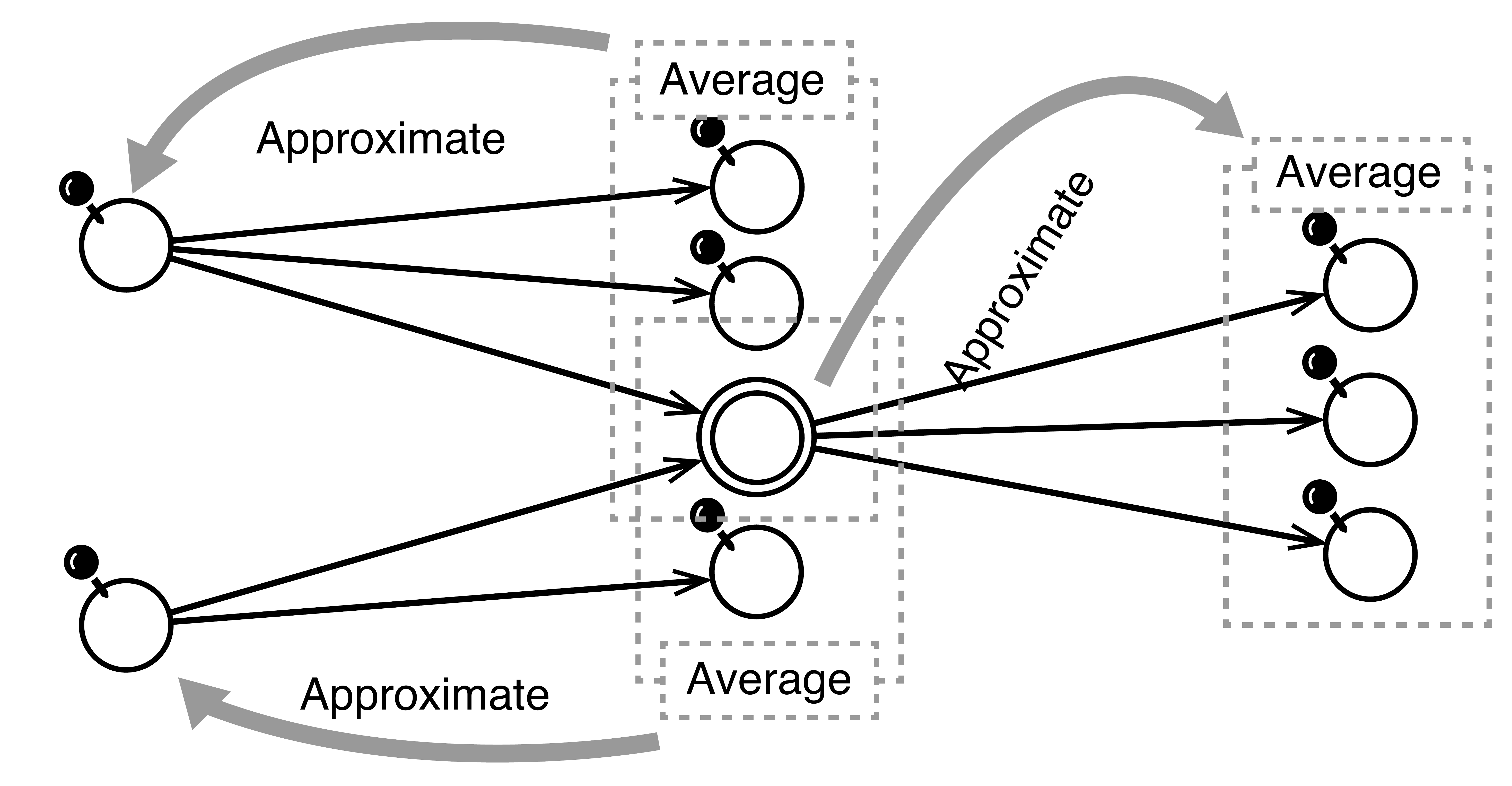}}
    \caption{Gradient descend focusing on a method name}
    \label{fig:optimization_of_a_embedding}
\end{figure}

Fig. \ref{fig:optimization_of_a_embedding} shows an example of the gradient descent focusing on a method name $m$ (shown as the double lined circle at the center) according to the partial derivative of the first term of $L$.
Method name $m$ not only moves to the average of the callees (rightmost box)
but also moves in the direction that the averages of $m$ and its siblings which are called from the same method (two boxes at the center) approximate the corresponding caller.
Thanks to this, embeddings of names of the method having no callees can have meaningful embedding based on how the methods are used.


In negative sampling, for each method $m$, a fixed set of methods that are not directly connected to $m$ is randomly taken as negative samples, then $m$ moves slightly toward a vector which is orthogonal to each negative samples.

\subsection{Recommendation phase} \label{subsec:build_call_graph}
The recommendation phase calculates the embedding of the method body given as a query from a user and then presents method names whose embeddings are similar to the query in order of the similarity.
In the calculation of the embedding of the query, the embeddings for the method names appearing in the body of the query are searched from the database, and the found embeddings are averaged.
The cosine similarity is used as the similarity measure for embeddings.

\section{Evaluation Experiment}
\begin{figure*}[tbp]
  \centerline{\includegraphics[width=2\columnwidth,pagebox=cropbox,clip]{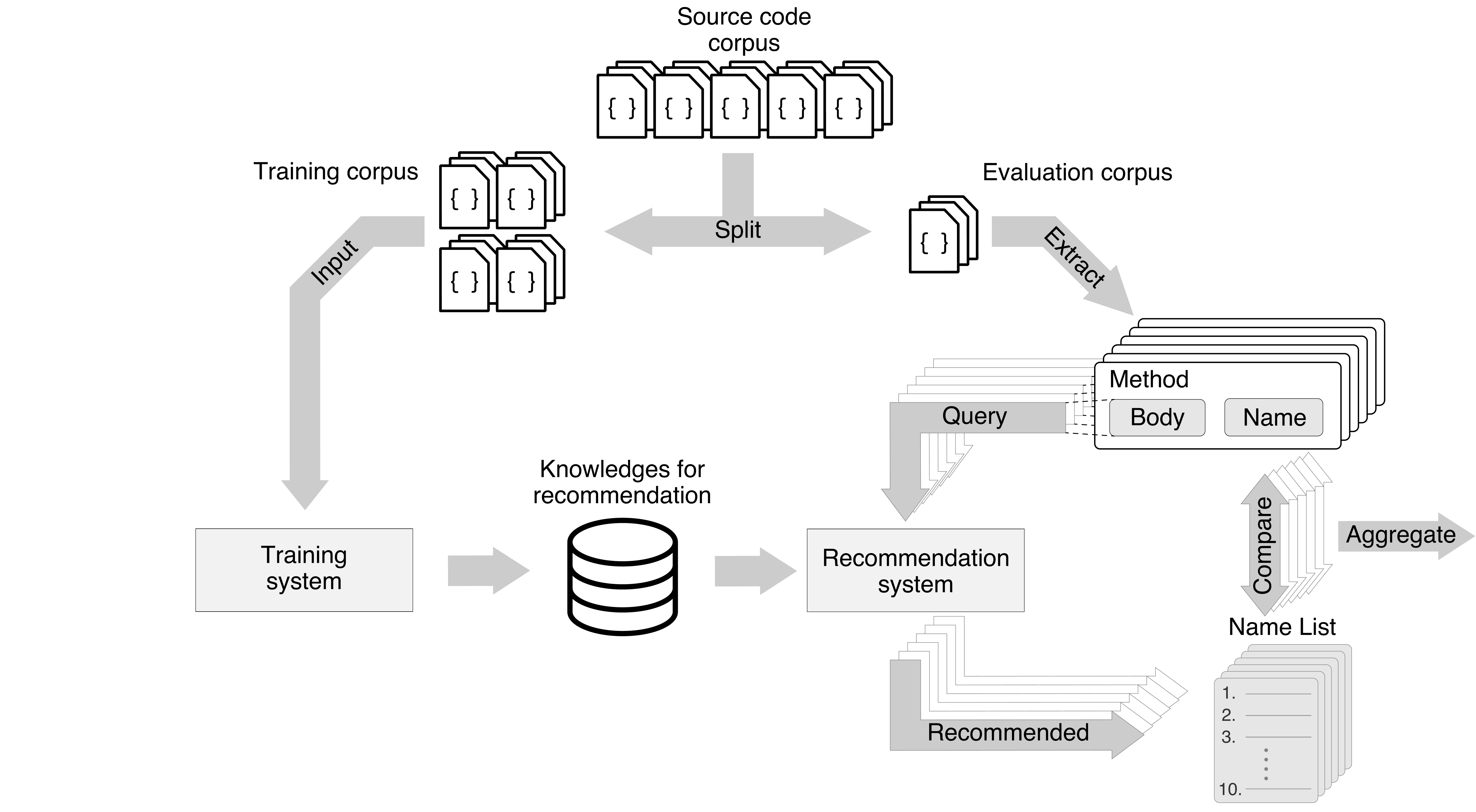}}
  \caption{Overview of the evaluation experiment}
  \label{fig:experiment_overview}
\end{figure*}

This section describes an evaluation experiment and its result.
The purpose of the experiment is to evaluate whether the proposed approach can more correctly recommend method names than the existing approach in the difficult situation, that is, a target method is not a getter or a setter and words used in the correct method name are unavailable in their bodies.
For quantitative comparison, this experiment measures how many identifiers in a proven code corpus the recommendation systems correctly recommend.
Then, we exploratorily investigate factors in the results of the evaluation.


In this experiment, the correctness of names recommended by the existing technique (Allamanis16)\cite{allamanis2016convolutional} and
the correctness of names recommended by the proposed technique are compared.
The reasons for choosing Allamanis16 as comparison target are as follows: 1) both the techniques are applicable before the target method is used, and 2) as far as we know, Allamanis16 is the state-of-the-art technique from the viewpoint of correctness.
The procedure to calculate the correctness is shown in Fig. \ref{fig:experiment_overview}.
Both the techniques consist of training and recommendation subsystems.
According to cross validation, a code corpus is split into a training corpus and a test corpus.
The whole training corpus is input to a training system.
For each method in the test corpus, recommendation result (i.e. candidate list of method names) for the method body are compared to the true method name, then the comparison results are aggregated.

The code corpus used for evaluating Allamanis16\cite{allamanis2016convolutional} is employed in this experiment.
The corpus consists of snapshots of 20 famous Java repositories on GitHub.
These famous products have been continuously developed for a long time so that methods in the projects are expected to be carefully named in general.

The correctness criteria are the same as \cite{kashiwabara2014recommending}: the top-10 candidates for a method contain at least one correct verb (or noun) or not.  The results are aggregated for three method categories, which are described below.

Methods are classified into three categories according to the richness of hints for naming contained in each body.
The category with the richest hints consists of getters and setters, namely,  the verb part of the method is ``get'' or ``set''.
Most of getters and setters strongly tend to have a boilerplate body, and ordinary development environments have a function to generate getters and setters for fields, so that developers are rarely bothered with naming getters and setters.
Both the second and third categories consist of methods other than getters and setters.
The second and third category are distinguished by the query to the recommendation subsystems contains true verbs (or nouns) or not.
Methods in the third category are considered the most difficult to be named.
The correctness is aggregated for each category.
Higher correctness of recommendation for the category with the least hints indicates an ability to support developers for naming.
Note that the inputs are system dependent; the input to the proposed system is a set of names of the callees, and the input to Allamanis16 is all tokens that compose the method body. For example, if a correct noun appears only in local variables within the body of a method that is neither a getter nor a setter, this method is included in the third category for evaluation of the proposed technique, and in the second category for evaluation of Allamanis16.

Please note that the following points to interpret the result of the experiment.\begin{itemize}
  \item Numbers of extracted methods differ depending on the systems,
	since the method extraction routine of Allamanis16 is inseparable and different from the corresponding one of the proposed system. (The proposed approach extract more methods.)
  \item Methods with no callees are removed for the evaluation of the proposed system,
	since the proposed system can not extract query form such methods.
  \item Combining the above two effects, the whole corpus of the proposed technique is 6\% larger than one of Allamanis16, and the number of getters and setters is almost 40\% larger for Allamanis16.
  \item Methods which contain no verbs (nouns) are excluded from the evaluation on verbs (nouns).
  \item Criteria to distinguish the second and third category is dependent on the systems as mentioned above.
\end{itemize}

The following are the set up for the experiments.
As a data cleansing, Java source files undesirable for the experiment are removed according to the following criteria.
\begin{itemize}
    \item Source files composed only from serially numbered methods
    such as get0(), get1() ..., get100().

    These files are automatically generated so that these methods are not named by developers.
    \item The word ``test'' is contained in the package name since almost all classes in these files are unit test code.
\end{itemize}
Then, the code corpus is split into 5 partitions file by file for 5-fold cross validation.
Part of speech of the words in the methods is determined by POSSE\cite{gupta2013part}.

The parameters for the proposed approach is as follows.  The number of dimension of embeddings is 100.  As to SGD, it loops 5000 times, a mini batch consists of 200 methods, and 10 negative samples are selected for each method. The learning rate is initially 0.75 and decreases 4\% after every iteration.

The parameters for Allamanis16 are the default values described in the official source code\cite{allamanis2016convolutional-implementation}, since these values should be the same as values used in the evaluation experiment in \cite{allamanis2016convolutional}.

\subsection{Result of the experiment}

Table \ref{tbl:result_for_verbs} and Table \ref{tbl:result_for_nouns} show recommendation correctness of the two approaches on verbs and nouns respectively.
Left and right half of the tables correspond to the proposed approach and Allamanis16, and both the halves are split into the three categories of the richness of the hints.
The bottom lines show the results for the whole corpus, and the middle lines show the results for each the partition of the cross validation.
In the results for the whole corpus, higher scores in the same category are bolded.

\begin{table*}[tb]
 \caption{Correctness of recommendation for verb part}
 \label{tbl:result_for_verbs}
 \centering
 \scriptsize
 \renewcommand\tabcolsep{2pt}
  \begin{tabular}{|l||l|l|l||l|l|l|}
  \hline
        & \multicolumn{3}{c||}{Mercem (proposed approach) } & \multicolumn{3}{c|}{Allamanis16} \\ \hline \hline
     & \multirow{2}{*}{getter/setter} & \multicolumn{2}{c||}{methods except for getter/setter}                                                                         & \multirow{2}{*}{getter/setter} & \multicolumn{2}{c|}{methods except for getter/setter}                                                             \\ \cline{3-4} \cline{6-7}
 \cline{3-4} \cline{6-7}
    &                                        & contains any of correct verbs & contains no correct verbs &                                & contains any of correct verbs & contains no correct verbs \\ \hline \hline
  part1                             & 69.45\% (3137 / 4517)                  & 66.53\% (3531 / 5307)                                 & 23.19\% (1531 / 6602)                                          & 89.68\% (5910 / 6590)          & 43.40\% (1860 / 4286)                       & 16.24\% (827 / 5092)                                 \\ \hline
  part2                             & 65.96\% (3384 / 5130)                  & 66.48\% (4011 / 6033)                                 & 25.19\% (1631 / 6476)                                          & 84.58\% (5723 / 6766)          & 45.18\% (1991 / 4407)                       & 17.80\% (884 / 4967)                                 \\ \hline
  part3                             & 68.22\% (3480 / 5101)                  & 65.25\% (3801 / 5825)                                 & 24.39\% (1744 / 7150)                                          & 87.55\% (6507 / 7432)          & 46.61\% (2337 / 5014)                       & 17.01\% (917 / 5390)                                 \\ \hline
  part4                             & 68.02\% (3397 / 4994)                  & 65.70\% (3436 / 5230)                                 & 23.35\% (1546 / 6620)                                          & 87.83\% (5983 / 6812)          & 52.12\% (2030 / 3895)                       & 17.96\% (872 / 4854)                                 \\ \hline
  part5                             & 72.09\% (3399 / 4715)                  & 66.62\% (3908 / 5866)                                 & 26.01\% (1867 / 7178)                                          & 84.65\% (5278 / 6235)          & 50.40\% (2718 / 5393)                       & 17.73\% (975 / 5499)                                 \\ \hline \hline
    \multirow{3}{*}{Total}                   & \multirow{2}{*}{68.68\% (16797 / 24457)}                & \textbf{66.12\%} (18687 / 28261)                               & \textbf{24.45\%} (8319 / 34026)                                         & \multirow{2}{*}{\textbf{86.90\%} (29401 / 33835)}        & 47.56\% (10936 / 22995)                     & 17.34\% (4475 / 25802)                               \\ \cline{3-4} \cline{6-7}
 &                                        & \multicolumn{2}{c||}{\textbf{43.36\%} (27006 / 62287)}                                                                           &                                & \multicolumn{2}{c|}{31.58\% (15411 / 48797)}                                                       \\ \cline{2-7}
               & \multicolumn{3}{c||}{50.50\% (43803 / 86744)}                                                                                                                    & \multicolumn{3}{c|}{\textbf{54.23\%} (44812 / 82632)}                                                                                        \\ \hline
  \end{tabular}
\end{table*}

\begin{table*}[tb]
 \caption{Correctness of recommendation for noun part}
 \label{tbl:result_for_nouns}
 \centering
 \scriptsize
 \renewcommand\tabcolsep{2pt}
  \begin{tabular}{|l||l|l|l||l|l|l|}
  \hline
        & \multicolumn{3}{c||}{Mercem (proposed approach)} & \multicolumn{3}{c|}{Allamanis16} \\ \hline \hline
     & \multirow{2}{*}{getter/setter} & \multicolumn{2}{c||}{methods except for getter/setter}                                                                         & \multirow{2}{*}{getter/setter} & \multicolumn{2}{c|}{methods except for getter/setter}                                                             \\ \cline{3-4} \cline{6-7}
 \cline{3-4} \cline{6-7}
    &                                        & contains any of correct nouns & contains no correct nouns &                                & contains any of correct nouns & contains no correct nouns \\ \hline \hline
  part1                             & 46.70\% (1951 / 4178)                  & 66.13\% (1355 / 2049)                               & 42.10\% (4074 / 9676)                                         & 81.34\% (5141 / 6320)                  & 71.33\% (5234 / 7338)                       & 21.31\% (681 / 3195)                                  \\ \hline
  part2                             & 43.88\% (2095 / 4774)                  & 65.37\% (1376 / 2105)                               & 42.77\% (4187 / 9790)                                         & 81.12\% (5265 / 6490)                  & 67.86\% (4874 / 7182)                       & 20.47\% (701 / 3425)                                  \\ \hline
  part3                             & 42.48\% (2023 / 4762)                  & 63.38\% (1547 / 2441)                               & 41.30\% (4322 / 10466)                                        & 81.52\% (5838 / 7161)                  & 73.64\% (5989 / 8133)                       & 17.32\% (612 / 3534)                                  \\ \hline
  part4                             & 44.66\% (2082 / 4662)                  & 72.69\% (1376 / 1893)                               & 41.95\% (3814 / 9092)                                         & 81.82\% (5361 / 6552)                  & 73.59\% (4868 / 6615)                       & 17.07\% (500 / 2929)                                  \\ \hline
  part5                             & 48.59\% (2135 / 4394)                  & 70.39\% (1745 / 2479)                               & 41.45\% (4356 / 10509)                                        & 82.62\% (4926 / 5962)                  & 74.32\% (6126 / 8243)                       & 22.11\% (841 / 3803)                                  \\ \hline \hline
  \multirow{3}{*}{Total}            & \multirow{2}{*}{45.17\% (10286 / 22770)}                & 67.47\% (7399 / 10967)                              & \textbf{41.90\%} (20753 / 49533)                                       & \multirow{2}{*}{\textbf{81.67\%} (26531 / 32485)}                & \textbf{72.22\%} (27091 / 37511)                     & 19.75\% (3335 / 16886)                                \\ \cline{3-4} \cline{6-7}
   &                                        & \multicolumn{2}{c||}{46.53\% (28152 / 60500)}                                                                        &                                        & \multicolumn{2}{c|}{\textbf{55.93\%} (30426 / 54397)}                                                        \\ \cline{2-7}
                 & \multicolumn{3}{c||}{46.16\% (38438 / 83270)}                                                                                                                 & \multicolumn{3}{c|}{\textbf{65.56\%} (56957 / 86882)}                                                                                                 \\ \hline
  \end{tabular}
\end{table*}


In the third category with the least hints, the proposed approach outperformed the other approach for both verbs and nouns.
From these result, we can conclude that the proposed approach should have a higher ability to support developers for naming.
For the verbs, the proposed technique shows higher correctness even in the second category, which indicates that the technique is effective for recommending a wide range of verbs.
On the other hand,  Allamanis16 also achieved higher correctness for nouns in the second category and the average of the second and third categories, which indicates that Allamanis16 also can effectively recommend many nouns.
However, the correctness of the proposed approach in the third category is more than twice that of Allamanis16, and the number of methods in the third category for the proposed technique is about 3 times as large as that of Allamanis16.
Therefore, no change is necessary to the conclusion that the proposed approach has a higher ability to support developers.

\subsection{Exploratory Investigation}

\begin{figure*}[tb]
  \begin{center}
    \begin{tabular}{c}
      \begin{minipage}{0.66\columnwidth}
        \begin{center}
          \includegraphics[width=\linewidth,clip]{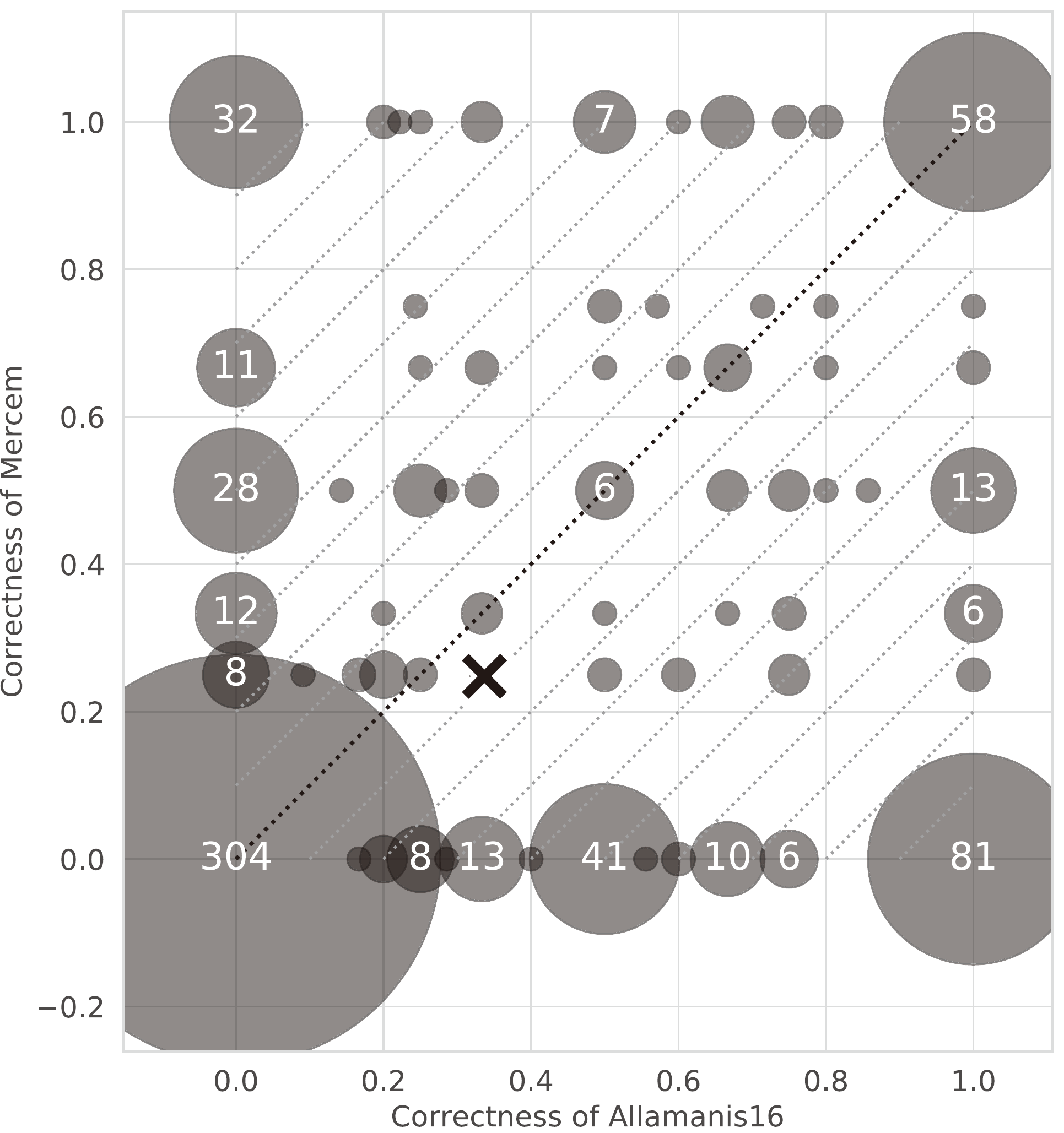}
          \caption{Comparison of correctness for verbs which appears 1-5 times}
          \label{fig:verb_correctness_1}
        \end{center}
      \end{minipage}

      \begin{minipage}{0.66\columnwidth}
        \begin{center}
          \includegraphics[width=\linewidth,clip]{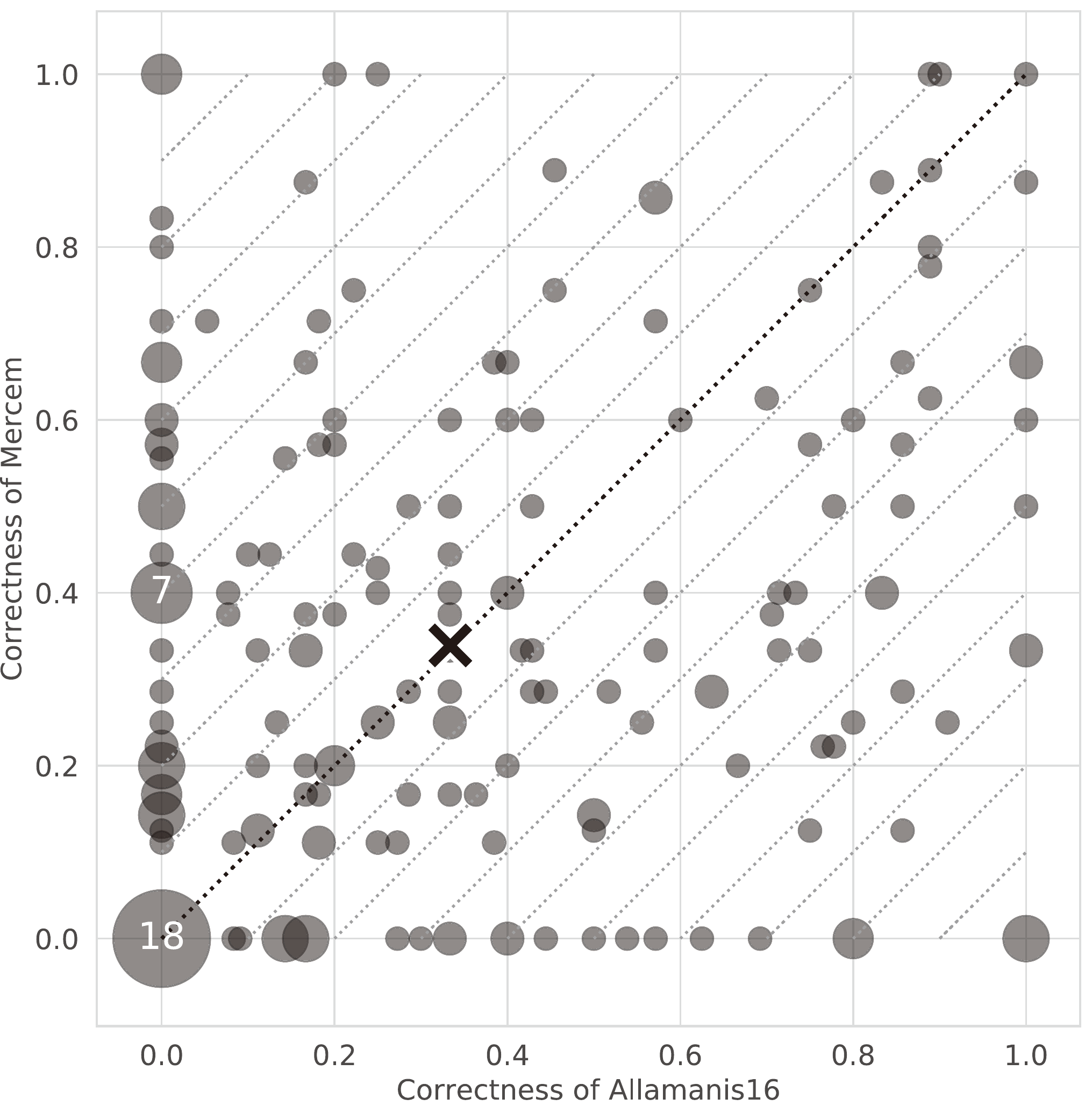}
          \caption{Comparison of correctness for verbs which appears 6-10 times}
          \label{fig:verb_correctness_2}
        \end{center}
      \end{minipage}

      \begin{minipage}{0.66\columnwidth}
        \begin{center}
          \includegraphics[width=\linewidth,clip]{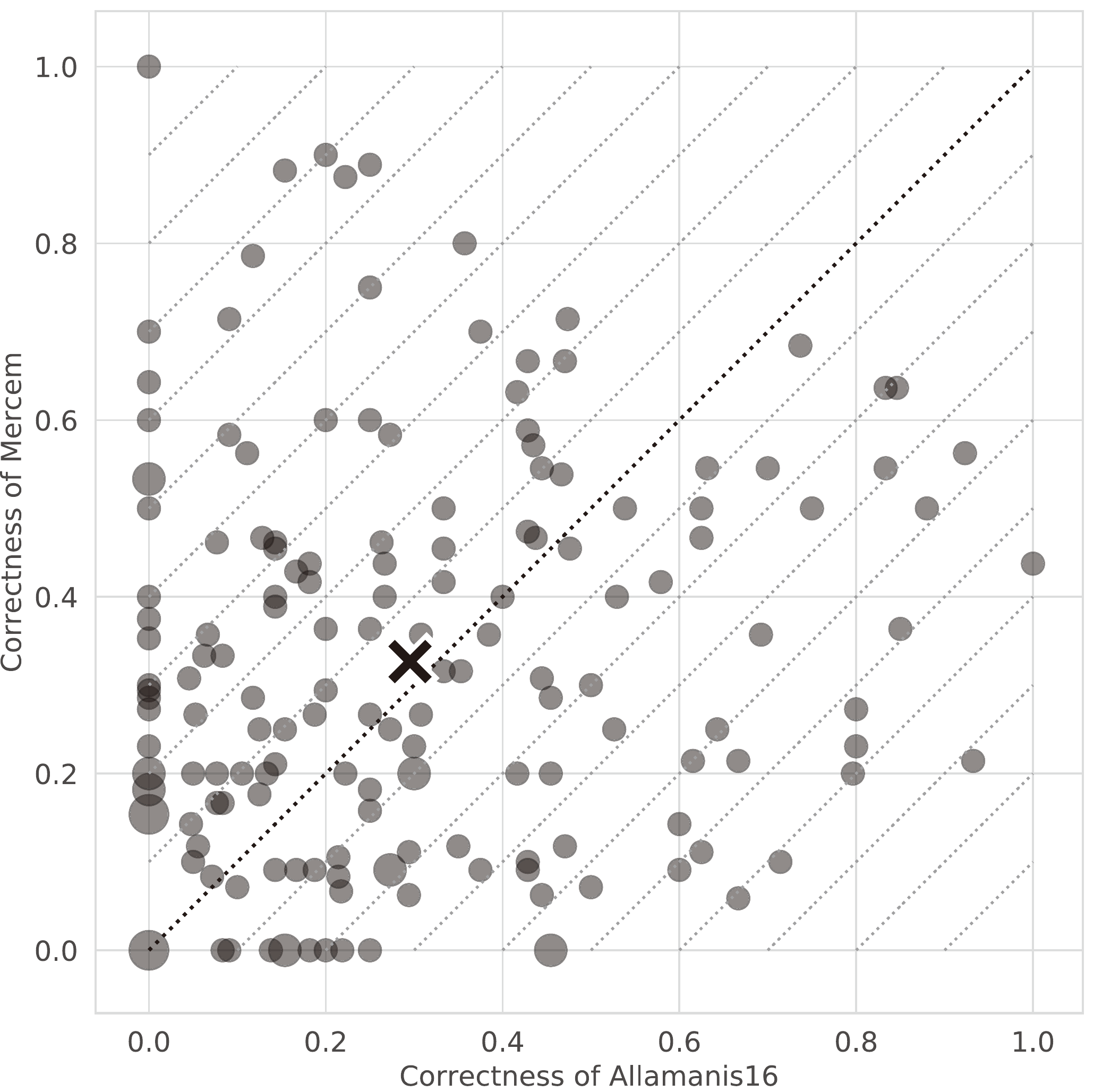}
          \caption{Comparison of correctness for verbs which appears 11-20 times}
          \label{fig:verb_correctness_3}
        \end{center}
      \end{minipage}
    \end{tabular}
  \end{center}

\end{figure*}

\begin{figure}[tb]
  \begin{center}
    \includegraphics[width=0.66\columnwidth,clip]{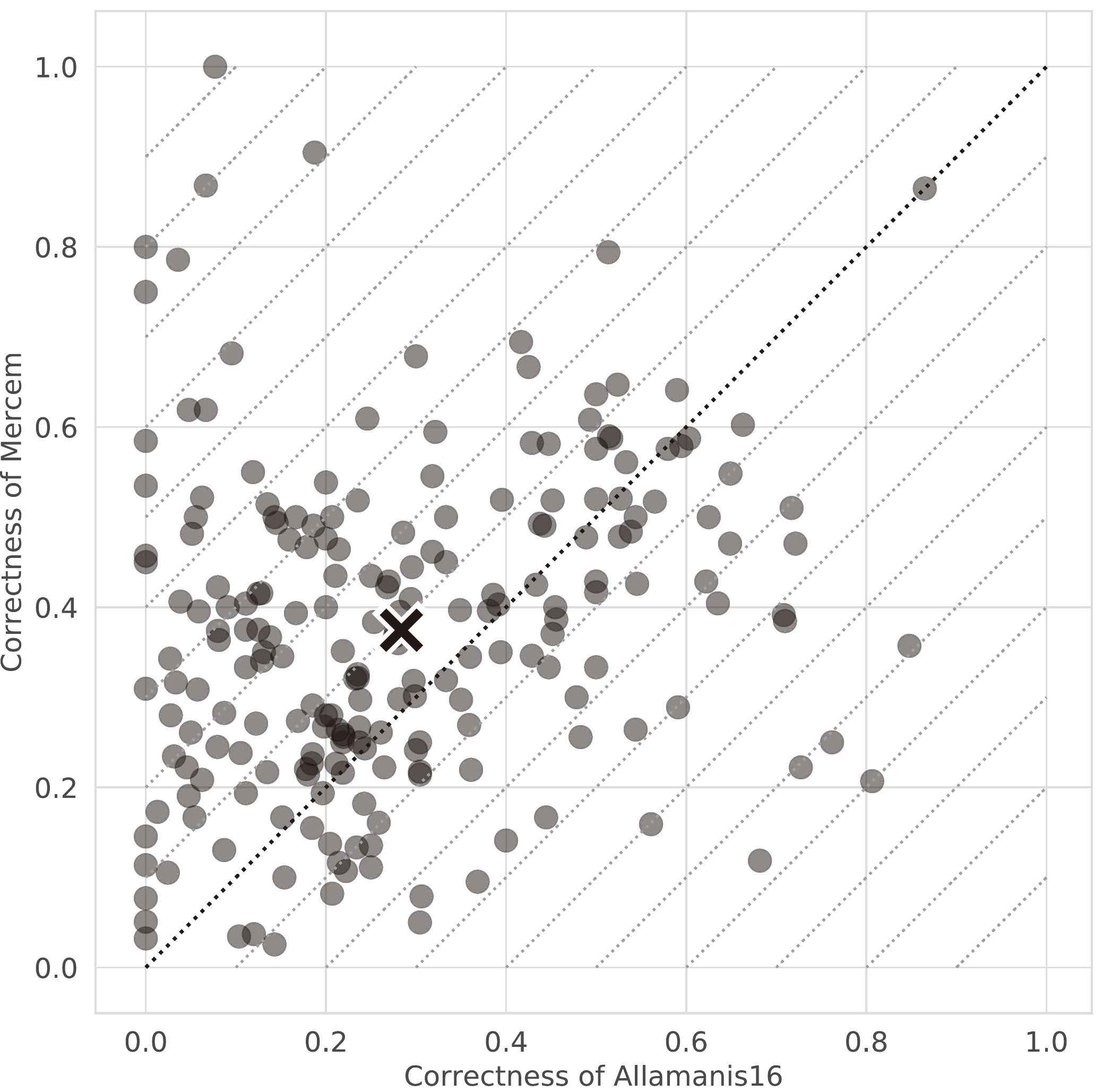}
    \caption{Comparison of correctness for verbs which appears 21-100 times}
    \label{fig:verb_correctness_4}
  \end{center}
\end{figure}

\begin{figure}[tb]
  \begin{center}
    \includegraphics[width=0.66\columnwidth,clip]{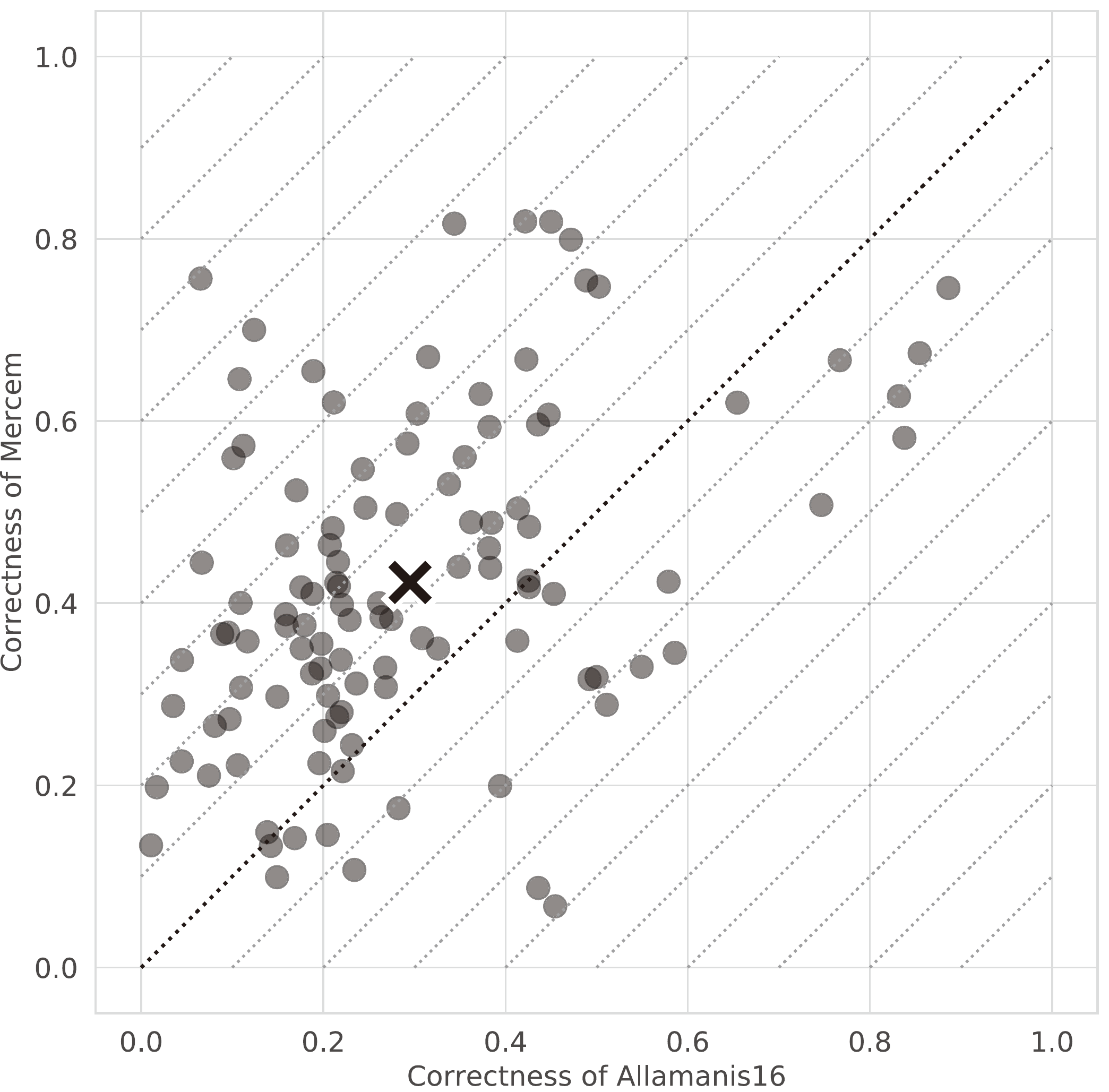}
    \caption{Comparison of correctness for verbs which appears 101+ times}
    \label{fig:verb_correctness_5}
  \end{center}
\end{figure}

In order to find out the cause of a significant advantage of the proposed approach for verb recommendation, the correctness of the two approaches is investigated according to the frequency of each verb in the method names.
Fig. \ref{fig:verb_correctness_1} to \ref{fig:verb_correctness_5} show the comparison of the correctness for each word grouped by the frequency of the verbs.
Each figure corresponds to the groups of verbs that appear 1-5, 6-10, 11-20, 21-100 or 100+ times respectively.
Vertical and horizontal axes mean the correctness of the proposed approach and Allamanis16 respectively.
One of the smallest circles in the graph represents that the proposed approach and Allamanis16 have recorded certain correctness for one verb.
Larger circles mean the records of verbs, whose number is proportional to the area of the circle.
The numbers of verbs are also shown as characters in the circles whose area is larger than 5.
The cross signs in the figures are the center of gravity of the circles.
The closer the circles or the cross signs to the upper left corner, the more correct the proposed approach is over Allamanis16.
Note that in Fig. \ref{fig:verb_correctness_1}, many verbs share the same position, as the correctness is limited to the rational numbers with a denominator of 1 to 5.


In the groups of highly frequent verbs (frequency is 21-100 (Fig. \ref{fig:verb_correctness_4}) or 100+ (Fig. \ref{fig:verb_correctness_5})), the correctness of the proposed approach is significantly higher for many verbs.
Because verbs in these groups are used in many methods, it is possible that this result has an impact on the correctness for the whole corpus, and thus caused the advantage of the proposed technique.

The trend that the correctness of the proposed approach is higher for the highly frequent verb can be confirmed by tracking the center of gravity in multiple figures.
From Fig. \ref{fig:verb_correctness_1} to Fig. \ref{fig:verb_correctness_5}, the center of gravity moves monotonously upward.

In the group of the most infrequent verbs (Fig. 5), Allamanis16 is more correct.  However, in this group, the correctness of both approaches was low, and the correctness of both approaches was zero for about half of the verbs.

\section{Discussion}
Since the output of the proposed technique was more correct than the existing technique in the most difficult situations, the proposed technique is expected to support developers to name methods more effectively.
Additionally, this result indicates that the approach for obtaining embeddings from the call graph has an ability to express a part of the function and/or the role of methods.

For the second category, where the target is not getter or setter and the input contains the words in the correct method name, the result was divided.
Therefore, the selective use of the proposed technique and the existing technique could be effective for supporting naming.

Next, let us consider the result of the exploratory investigation.
For the highly frequent verbs, the correctness of the proposed technique tends to be high.
By contrast, for the less frequent verbs, the correctness of both techniques was low, and several verbs were never correctly suggested by the techniques.
It is thought that this result has led to the omission of information related to the infrequent words, since both the graph embedding and the neural network are based on statistical approximations.

From the above discussion, let us consider how to improve the recommendation.
The proposed technique and Allamanis16 perform well in different situations so that ensembling of the two techniques could improve the recommendation correctness.
On the other hand, different approaches are required for infrequent words since the correctness of both the techniques is low for these words.
The combination with rule-based approaches, such as Kashiwabara's association rule based approach\cite{kashiwabara2014recommending}, might be promising for improving the correctness for infrequent words, since it is known that rule-based techniques can capture infrequent relationships if there are strong correlations.

\subsection{Threats to Validity}

\subsubsection*{Internal Validity}
With regard to the first category (getters and setters) and the category that merged the second and third, the two techniques are compared fairly because the criteria for the separation is the same.  On the other hand, there are concerns about comparing the techniques in the second or third category,  because the separation criteria differ depending on the techniques.  However, we believe that the comparison satisfies one aspect of fairness since the separation is based on the data that each of the two techniques requires as input.  In the future, it is necessary to consider comparative experiments that satisfy another perspective of fairness.

\subsubsection*{External Validity}
There is a concern about the generality of the input code corpus.
Since the source code in the corpus is widely used and contributed by many developers, there is no problem to employ the method names in the corpus as the ground truth.
On the other hand, since the code in the corpus is biased towards high quality, it is unclear that this result can be generalized to low-quality code.
However, since there are few alternatives to the API sets for implementing a certain function, the set of callee methods is stable regardless of the code quality.
Therefore, the correctness of the proposed technique is expected to be robust to the code quality.
By contrast, the correctness of Allamanis16 depends on the abilities to pick up words from the method body or to estimate words from code idioms used in the method.
So that when code quality is low, for example, parameter names and local variable names are wrong, or code idioms are not properly used, the accuracy of Allamanis16 could tend to be low.

\section{Conclusion and future work}
%
%

This paper proposed an approach to recommend method names to a software developer using method embeddings obtained from caller-callee relationships.
The evaluation experiment confirmed that the correctness of the proposed approach is higher than the state of the art approach especially in the situation in which the naming is difficult.
An embedding of a callee method is approximated to the average of embeddings of callee methods, thereby embeddings successfully express functions of the methods.


We have not yet confirmed whether the proposed approach can support the developer's naming in real development situations, although the proposed approach is expected to recommend more correct names in the real situation.
In the future, evaluation experiments in situations closer to real development situations are needed.
At the same time, it is required to improve the correctness of the proposed approach.
It is possible to improve the correctness by refining the calculation the embeddings from caller-callee relationships or combining with the existing approaches\cite{allamanis2016convolutional,kashiwabara2014recommending}.





\end{document}